# Unconventional S-orbital state of Tb and cooperative Ru(4d)-Tb(4f) spin-ordering in strongly correlated 4*d*-4*f* system, Ba$_3$TbRu$_2$O$_9$

E. Kushwaha,[1] G. Roy,[1] A. M. dos Santos,[2] M. Kumar,[1] S. Ghosh,[1] T. Heitmann,[3] and T. Basu[1, *]

[1] Department of Sciences and Humanities, Rajiv Gandhi Institute of Petroleum Technology, Jais, Amethi, 229304, Uttar Pradesh, 229305, India

[2] Neutron Scattering Division, Oak Ridge National Lab, Oak Ridge, TN 37831, USA

[3] The Missouri Research Reactor, University of Missouri, Columbia, Missouri 65211, USA

*Corresponding Author: tathamay.basu@rgipt.ac.in

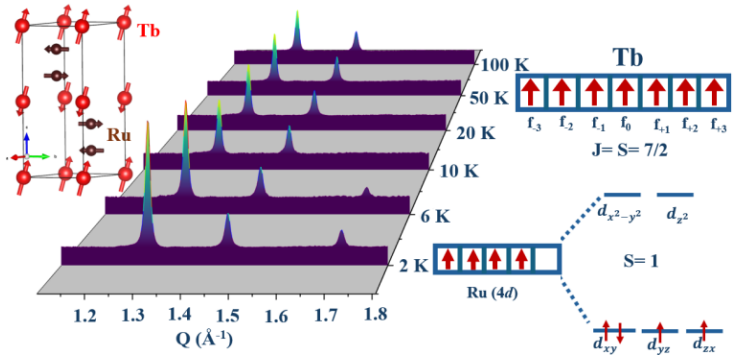

*Concept:* We resolve the intriguing spin-structure of a 4d-4f system, Ba$_3$TbRu$_2$O$_9$, which shows a cooperative spin-ordering of both Ru(4d) and Tb(4f), not commonly observed in d-f coupled systems. We demonstrate an unconventional s-orbital state of terbium (4f-ion), exhibiting zero orbital moment (L=0), and an S=1 ground state of ruthenium with full spin moment, in contrast to all other members in this family.

## Abstract

The 6H-perovskite Ba$_3$RRu$_2$O$_9$ (R = rare-earth), composed of Ru$_2$O$_9$ dimers connected through RO$_6$ octahedra, exhibits an intriguing variety of magnetic ground states – ranging from non-magnetic to ferromagnetic and antiferromagnetic, depending on the specific R ion. In this study, we investigate the compound Ba$_3$TbRu$_2$O$_9$ using magnetic susceptibility measurements and time-of-flight neutron diffraction experiments. Our combined bulk and microscopic analyses reveal that the Tb$^{4+}$ (4$f^{\,7}$) electronic configuration results in an s-like state with an orbital moment **L** = 0 and spin-only value of **S** = 7/2, and Ru$^{4+}$ exhibits a spin-only value of **S** = 1 despite the presence of strong spin-lattice coupling in this compound, representing a sharp contrast to other reported members of this family. A cooperative 4*d*–4*f* spin ordering is observed below the Néel temperature (T$_N$ ~ 9.5 K), indicating strong Ru(4*d*)–Tb(4*f*) correlations in the system. The Tb-moments order antiferromagnetically in the bc-plane, whereas the Ru-moments are aligned antiferromagnetically along the b-axis. Furthermore, a collinear antiferromagnetic arrangement of spins is observed within the Ru$_2$O$_9$ dimers throughout the structure, unlike other reported members of this series (e.g., Ho and Nd).

1. ## Introduction

The 6H-perovskite system, A$_3$MM′$_2$O$_9$ (where A = Ca, Sr, Ba; M= Li, Bi, 3*d*-transition metal, or lanthanide metal ion (Ln); M′= 4*d*/5*d*-transition metal ion), exhibits a variety of exotic magnetic ground states arising from strong metal–metal (M-M′) electronic correlations and high magnetic frustration, which leads to various quantum phenomena, such as quantum spin liquid behavior, molecular-type magnetism or spin-driven ferroelectricity [1–8]. The Ba$_3$LnRu$_2$O$_9$ system has attracted significant attention due to the strong Ru(4$d$)-Ln(4$f$) interactions, strong spin-orbit coupling (SOC) and large crystal electric field (CEF) of the Ru (4$d$)-orbitals, and rare-earth anisotropy [9–12]. In these 6H perovskite systems, the Ru$_2$O$_9$ dimers, consisting of two face-sharing distorted RuO$_6$-octahedra, are connected through corner-sharing regular LnO$_6$ octahedra (Fig.1a and 1b). For non-magnetic lanthanide ions (Ln= Y, La, Eu, Lu), the compound behaves like a magnetic dimer, exhibiting a broad hump in temperature-dependent magnetic susceptibility [3,10]. Conversely, the presence of a magnetic rare-earth introduces long-range magnetic ordering. Interestingly, the Nd member of this family shows ferromagnetic (FM) ordering at 24 K, in sharp contrast to all other rare earth (Sm, Gd, Tb, Ho, Dy, Er) members which exhibit a long-range antiferromagnetic (AFM) ordering at 9–12 K [4,5,11–14]. Recently, systems with heavy rare-earths (R= Ho, Tb,) were characterized as magnetodielectric and showed potential as hosts of multiferroicity [4,11]. This intriguing magnetism and spin-driven ferroelectricity is well established for Ba$_3$HoRu$_2$O$_9$ [6].

It is clear that the lanthanide ion plays a decisive role in establishing the magnetic ground state in this family. The lanthanide-contraction across the series modifies both Ln-O-Ru bond angles, and the Ru-Ru distances, which in turn influence the magnetic exchange interactions. For example, in Ba$_3$YRu$_2$O$_9$ the ruthenium exhibits a **S** = 1/2 ground state due to strong metal-metal bonding [3]. Replacing yttrium with the larger lanthanum modifies the Ru-Ru distance, which results in **S** = 3/2 orbital-selective Mott ground state for Ba$_3$LaRu$_2$O$_9$ arising from different degrees of Ru–Ru hybridization [10]. Conversely, for Ba$_3$NdRu$_2$O$_9$, the 4$f$ orbital of Nd is less localized and closer to the Fermi level compared to those of other heavy rare-earth ions in this family. This might result in the unique observation of ferromagnetic ordering of the Nd member [11,12]. In most $d$–$f$ correlated systems, the d-orbital ions magnetically order at high temperature due to their extended bonding character, while the, more shielded, $f$-orbitals magnetically order at lower temperature. In contrast with this observation, cooperative Ru(4$d$)-Ho (4 $f$) spin ordering was already observed in Ba$_3$HoRu$_2$O$_9$ below 50 K [5]. Furthermore, a reduced Ru-moment compared to its spin-only (**S** = 2) value is reported for the Nd and Ho members in this series [5,12]. The exact ruthenium moment and nature of its ground state are also ambiguous in other R-members of this family. This overview highlights how the magnetic ground state of Ru and R ions varies significantly across the lanthanide series. In these oxides, the magnetic rare-earth ions are trivalent (R$^{3+}$) such that the Ru$_2$O$_9$ dimers have mixed valence Ru$^{4+}$/Ru$^{5+}$ states. The only exception is for Ba$_3$CeRu$_2$O$_9$ and Ba$_3$TbRu$_2$O$_9$, where Ce/Tb can adopt a tetravalent (Tb$^{4+}$) configuration, resulting in a single valence Ru$^{4+}$ state. A recent neutron scattering experiment confirmed that Ce$^{4+}$ has a non-magnetic ground state and Ru does not magnetically order down to 2 K [15]. An earlier study on Ba$_3$TbRu$_2$O$_9$ reported long-range AFM ordering at 9.5 K, proposing the magnetic ordering of the Tb moment, but with no contribution ascribed to ruthenium [14]. Therefore, a detailed magnetic structure describing both the Tb and Ru ground states remains absent from the literature.

The ruthenium ground state often varies significantly due to small changes in the environment (such as hybridization and lattice distortions), resulting from the competition between the crystal field effect and spin-orbit coupling. For example, in the Ba$_4$Ru$_3$O$_{10}$, which consists of Ru$_3$O$_{10}$ trimers, the central Ru$^{4+}$ ion exhibits a non-magnetic ground state (**S** = 0), while the ground state of the edge Ru$^{4+}$ ions in the trimer is magnetic (**S** ≠ 0) [16]. It has also been demonstrated that different RuO$_6$ octahedral distortions, even in the same structure, play a significant role in deciding magnetic ground state, yielding an intriguing magnetic structure in the same family [16,17]. The role of R-ions on local MO$_6$ octahedral-distortion and complex magnetism is also documented in many 3$d$–4$f$ systems, for example, in the canonical RMnO$_3$ perovskites [18]. Hence, the understanding of the ground state of Ru and Tb, and 4$d$–4$f$ correlation in the Ba$_3$TbRu$_2$O$_9$ compound is essential to understand the consequences of the subtle interplay of lattice-orbital-magnetic degrees of freedom in these compounds.

In this study, we have performed a detailed time-of-flight neutron diffraction experiment over a wide momentum transfer ($Q_{min}$ < 1 Å$^{-1}$), which allows us to resolve the spin-structure and magnetic ground state of both Tb and Ru. The results reveal that in $Ba_3TbRu_2O_9$, an unconventional spin-only ground state emerges, with Tb (orbital moment $L=0$, $J=S=7/2$) and spin-only moment of Ru ($S=2$, $L=0$).

## 2. Experimental details

The compound $Ba_3TbRu_2O_9$ was synthesized by solid- state-reaction using mixtures of high purity (>99.9%) precursors: $BaCO_3$, $RuO_2$, and $Tb_4O_7$ by mixing thoroughly using an agate mortar and pestle. The homogenized mixture was pressed into pellets and subjected to a series of calcination steps. The initial firing was carried out at 900 °C for 12 h, followed by successive heating at 1100 °C for 48 h, 1150 °C for 24 h, and a final sintering at 1180 °C for 12 h. After each heating cycle, the pellets were reground and repalletized to ensure homogeneity [14]. To check the phase purity powder X-ray diffraction was carried out using a Panalytical diffractometer fitted with a Cu-Kα source. Magnetization measurements as a function of temperature and magnetic field were performed using a Quantum Design Superconducting Quantum Interference Device (SQUID, Quantum Design). Time-of-flight (TOF) neutron diffraction data were collected at the SNAP beamline, diffractometer at Spallation Neutron Source (SNS) in Oak Ridge National Laboratory (ORNL), USA. The magnetic structure was solved using the FULLPROF and SARAH programs [19,20]. X-ray Photoelectron Spectroscopy (XPS) measurements were carried out using a Thermo Fisher Scientific K-Alpha spectrometer equipped with a monochromatic Al Kα source (1486.6 eV, 10 kV, 10 mA). Scanning Electron Microscopy (SEM) imaging was conducted using a JEOL JXA-8230 electron probe microanalyzer operated at an accelerating voltage of 20 kV. High-resolution high-angle annular dark-field scanning transmission electron microscopy (HAADF-STEM) was performed on a JEM-ARM200F instrument operating at 200 kV. For sample preparation, a small amount of finely grounded powder was dispersed in ethanol using ultrasonication for uniform mixing. A drop of the resulting suspension was then deposited onto a carbon-coated copper grid and allowed to dry at room temperature before imaging.

## 3. Results and discussion

### 3.1 Structural analysis and Magnetic susceptibility measurement

The Rietveld refinement of the X-ray diffraction pattern is shown in Fig. 1c, confirming the purity of the sample. A representation of the crystal structure obtained from the Rietveld refinement and the Tb-O-Ru-O-Tb super-exchange paths are shown in Fig. 1a and b. The Ru-Ru distance and Ru-O-Ru angle within the dimers is 2.56 Å and 79.26° respectively, indicating that the short Ru-Ru direct exchange interaction and the Ru-O-Ru super-exchange interactions are competing within the $Ru_2O_9$ dimers. However, these dimers are magnetically connected via a terbium. The Ru-O-Tb angle is 179.180°, implying a stronger AFM super-exchange path, over the Ru- O-Ru exchange interaction. Therefore, the Ru-O-Tb-O-Ru super exchange path is responsible for magnetic ordering.

To confirm the elemental composition and homogeneous distribution of constituent elements in the $Ba_3TbRu_2O_9$ powder sample, we performed scanning electron microscopy (SEM) with energy- dispersive X-ray spectroscopy (EDS) and high-resolution high-angle annular dark-field scanning transmission electron microscopy (HAADF-STEM). The SEM-EDS and HAADF-STEM images are shown in Fig. S1,

S2 and S3 in S.I. The SEM-EDS and HAADF-STEM elemental maps confirm a uniform spatial distribution of Ba, Ru, Tb, and O across the sample, confirming the homogeneity of the sample. The EDS results show that the atomic percentage ratio for Ba: Tb: Ru: O is 22.0: 7.2: 15.5: 60.6 (≈ 3.1: 1: 2.1: 8.4) which agree with the stoichiometry of the compound within the resolution limit of the instrument.

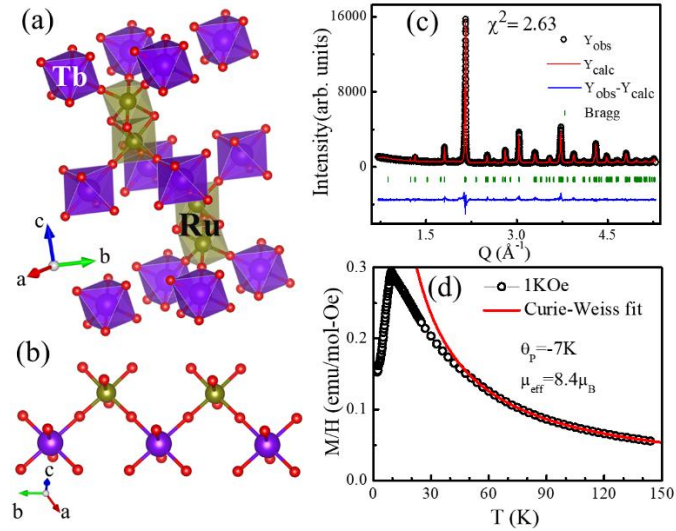

**Fig. 1 (a) Crystal structure of $Ba_3TbRu_2O_9$, (b) Tb-O-Ru-O-Tb ex- change path, (c) XRD Rietveld refinement at room-temperature, and (d) DC magnetic susceptibility as a function of temperature for a 1 kOe magnetic field.**

The temperature-dependent magnetic susceptibility under a 1 kOe magnetic field in zero-field-cooled conditions is shown in Fig. 1d. The results reveal antiferromagnetic (AFM) ordering below 9.5 K ($T_N$), consistent with an earlier report [14]. The Curie–Weiss fitting in the paramagnetic region yields an effective moment ($\mu_{eff}$) of 8.4 μB. This low value of μeff is not consistent with $Tb^{3+}$ (effective quantum number J = 6, $\mu_{eff}$ ~ 9.72 μB). Suggesting that terbium adopts a $Tb^{4+}$ valence with a spin configuration of S = 7/2, and L = 0 for a half-filled shell. This is consistent with the theoretical value of 7.94 μB for S = 7/2 is. The experimentally calculated effective magnetic moment per Ru atom is approximately 1.94 μB, based on the relation:

$$\mu_{eff}^2 = \mu_{Tb}^2 + 2\mu_{Ru}^2$$

To check the oxidation states of Tb and Ru we have performed XPS of the sample (shown in Fig.S4). The binding values of Tb $4d_{5/2}$ are 150.3 eV and $4d_{3/2}$ are 157.8 eV and for Ru $3p_{3/2}$ is 461.7 eV and $3p_{1/2}$ are 484.2 eV which are close to the values reported in the literature [21,22]. However, it should be noted that the binding energy difference between two oxidation states of Ru-atom is only nearly 1 eV, and such a small difference may also arise for the same oxidation state in different crystallographic environments [23,24]. However, if the valence state is a mixture of +4 and +5, we would observe two separate nearby peaks or a very broad peak, whereas here we observe a single peak due to single valence state as similarly observed in the literature [21-24].

## 3.2 Time-of-flight Neutron Diffraction and Magnetic Structure

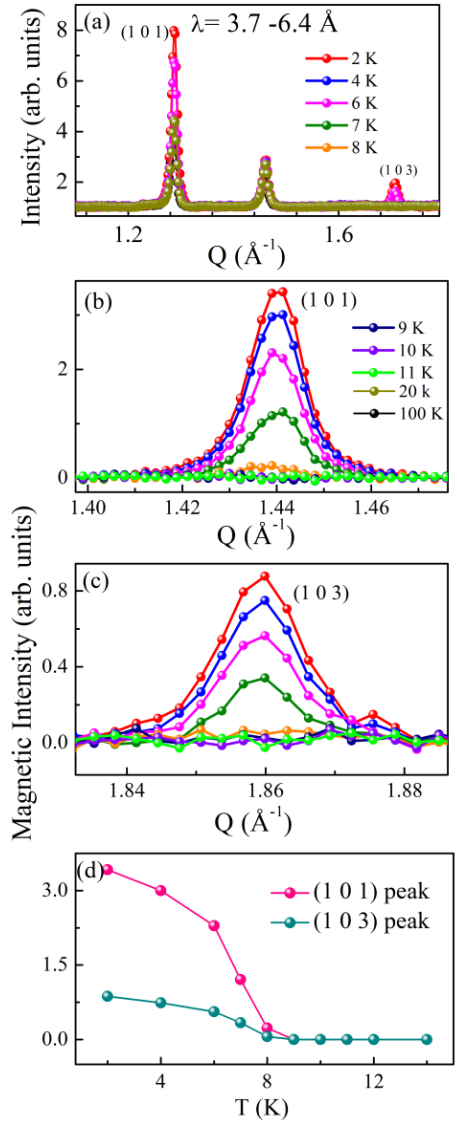

**Fig. 2** TOF neutron data of $Ba_3TbRu_2O_9$ at various temperatures (a) for λ=3.7-6.4, Magnetic intensity at various temperature for (b) (101) Bragg peak, (c) (103) Bragg peak and (d) Magnetic intensity vs temperature plot.

To solve the magnetic structure, we have performed temperature dependent (2, 4, 6, 7, 8, 9, 10, 12, 14, 16, 20, 50, and 100 K) time-of-flight neutron diffraction (ND) measurements and collected data over a wavelength band covering 3.7-6.5 Å, corresponding to a momentum transfer coverage between 1 Å$^{-1}$ and 2 Å$^{-1}$ (shown in Fig. 2a). A significant enhancement of the intensity of the (101) and (103) Bragg peaks (at Q = 1.44 Å$^{-1}$ and 1.86 Å$^{-1}$ respectively) is observed below $T_N$. No change in diffraction intensity is observed above $T_N$. The magnetic reflections of both Bragg peaks (101) and (103)) at various temperatures are shown in Fig. 2b and Fig. 2c. The magnetic intensities are calculated by subtracting the data collected at 2–11 K from the 100 K data which contains the purely nuclear contribution. The temperature dependence of the integrated magnetic intensity is shown in Fig. 2d, which represents the magnetic order parameter. The magnetic intensity increases with decreasing temperature below $T_N$ due to saturation of magnetic moments with lowering the temperature as usual. We did not observe any additional magnetic Bragg reflections at low-T in this Q-range. Therefore, we have further performed the experiments over a broader Q-range (for wavelength range 0.5–3.65 Å) at a few selected temperatures (100, 50, 20, and 2 K). Interestingly, 2 K diffraction profile shows a few extra peak intensities at Q = 1.44, 1.86, 2.19, 2.52, 3.30 and 3.53 Å$^{-1}$ corresponding to reflections (101), (103), (211), (105), (213) and (210) respectively compared to the high temperature diffraction profile above $T_N$, shown in Fig. 3a.

The neutron diffraction profile measured at 100 K in the paramagnetic region is well-fitted for the reported space group P6$_3$/mmc, confirming the phase purity of the sample (see Fig. 3b). All the additional magnetic reflections can be indexed using the propagation vector k = (0 0 0).

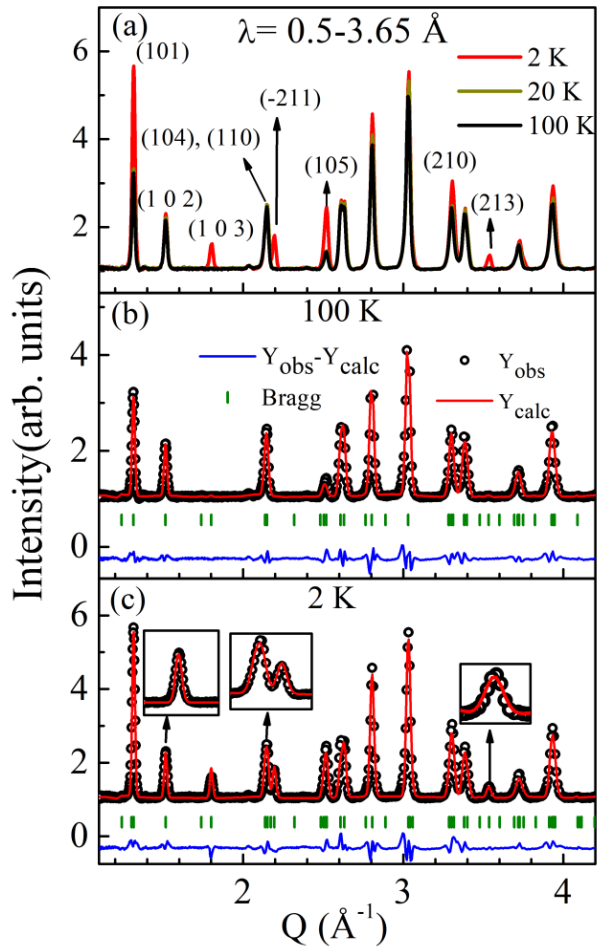

**Fig.3 TOF neutron data of Ba$_3$TbRu$_2$O$_9$ at various temperatures: (a) for λ = 0.5-6.5 Å, Rietveld refinement of TOF data (b) at 100 K and (c) at 2 K. Inset of (c) shows the enlarged view of the profile fitting. The magnetic and nuclear Bragg tick marks are in same position due to k = (0 0 0) wave vector.**

The Wyckoff positions of the magnetic Tb and Ru ions are 2a (0, 0, 0) and 4f (0.33333, 0.66667, 0.16223), respectively. To evaluate the possible magnetic structures consistent with the crystallographic symmetry, we performed an irreducible representational analysis using the SARAh program [19]. The resulting irreducible representations (IRs) and basis vectors (BVs) for the Ru and Tb sites are presented in Tables-S1 and S2. There are four irreducible representations for Tb and eight for Ru. The magnetic representation is given by:

$$\Gamma_{mag}(Tb) = 1\Gamma_3^1 + 1\Gamma_7^1 + 1\Gamma_9^2 + 1\Gamma_{11}^2$$

$$\Gamma_{mag}(Ru) = 1\Gamma_2^1 + 1\Gamma_3^1 + 1\Gamma_6^1 + 1\Gamma_7^1 + 1\Gamma_9^2 + +1\Gamma_{10}^2 + 1\Gamma_{11}^2 + 1\Gamma_{12}^2$$

Since neutrons interact with the magnetic moment component perpendicular to the momentum transfer (Q), the significant difference in intensity of the (1 0 1), (1 0 3), (1 0 4), (1 1 0), (1 0 5), (2 1 0) Bragg peak between 2 K and 100 K suggests that the magnetic moment should have a dominant contribution along b-axis. However, the presence of strong magnetic reflection for (2 1 0) and (-2 1 1) indicates there should be magnetic component along c-axis as well.

We could not achieve a suitable refined structure considering $\Gamma_2$, $\Gamma_6$, $\Gamma_{10}$ and, $\Gamma_{12}$, ruling out the possibilities of a non-magnetic terbium (because, only ruthenium contributes for these particular basis vectors, obtained from SARAh analysis). We have attempted to fit the data using $\Gamma_3$, $\Gamma_7$, $\Gamma_9$ and $\Gamma_{11}$, where both Tb and Ru moments contributes. For terbium, $\Gamma_9(\Psi_5)$ and $\Gamma_{11}(\Psi_5)$ have a magnetic moment in b- plane, $\Gamma_9(\Psi_4)$ and $\Gamma_{11}(\Psi_6)$ have magnetic moment in ab- plane and $\Gamma_3(\Psi_1)$ and $\Gamma_7(\Psi_2)$ have magnetic moment in c-axis (Table-S1). For ruthenium, $\Gamma_9(\Psi_5)$ and $\Gamma_{11}(\Psi_9)$ have magnetic moment along b-axis, $\Gamma_9(\Psi_6)$ and $\Gamma_{11}(\Psi_{10})$ have magnetic moment in ab- plane and $\Gamma_3(\Psi_2)$ and $\Gamma_7(\Psi_4)$ have magnetic moment in c-axis (Table-S2). The best fit to the data is achieved using a combination of $\Gamma_{11}(\Psi_5)$ for Ru and $\Gamma_7(\Psi_2)$, $\Gamma_{11}(\Psi_5)$ for Tb. The corresponding Shubnikov space group, which is the magnetic space group of the paramagnetic space group P6$_3$/mmc with k = (0, 0, 0), is P6'$_3$/m'm'c.

The Rietveld refinement of 2 K data is shown in Fig. 3c. The enlarged view of selected nuclear and magnetic peaks are shown in the inset of Fig. 3c and Fig.S5 in S.I. The fittings parameters are: $\chi^2$ = 9.8 Rp = 2.87, $R_{wp}$ = 4.27, $R_{exp}$ = 1.11, and magnetic R-factor, $R_{mag}$ = 5.14, which indicate the good fitting of the data. The magnetic structure obtained from this refinement is shown in Fig. 4a.

The results reveal that the Ru moments are ordered along the b-direction with a magnetic moment of 1.96 µB, and the Tb moments are ordered in the bc-plane (small canting of 15.9° with c-axis) with a magnetic moment of 6.18 µB. The moment associated with terbium and ruthenium are tabulated in Table-S3 in S.I. The fit of the magnetic structure was further checked by setting the Ru moment to zero and refining the Tb moments allowing it to achieve its maximum value; this, however, does not result in a good fit. The refined magnetic structure results in a terbium with 6.18 ± (0.04) µB (close to **S**=7/2) and a spin-only moment of ruthenium (**S**=1).

In most of the *d-f* coupled systems, the transition metal ion first orders at higher temperature, followed by the rare-earth ordering at lower temperature [25-30]. Interestingly, here we observe the rare cooperative ordering of Ru(4*d*) and Tb(4*f*) moments below $T_N$, which is attributed to strong 4*d*-4*f* coupling in this system.

The Ru-moments are collinearly arranged antiferromagnetically along the b-axis for the title compound (Fig. 4a). For Ba$_3$HoRu$_2$O$_9$ compound, the ruthenium network exhibits a canted AFM spin-structure within Ru$_2$O$_9$ dimers below magnetic ordering [5]. The Ru-spins are arranged ferromagnetically within Ru$_2$O$_9$ dimers for Nd-members, and ferromagnetic dimers are arranged antiferromagnetically [12]. The strong spin-orbit coupling and strong 4*d*-4*f* correlation is reported in heavy-rare-earth member Ba$_3$HoRu$_2$O$_9$. Surprisingly, being a heavy rare-earth member, the Tb possesses an **S**=7/2 ground state ($f^7$) with zero orbital moment, adopting an unusual +4 valence state, which is characterized as an s-state, more commonly observed for Gd$^{3+}$ ions, which have $f^7$ electronic configurations.

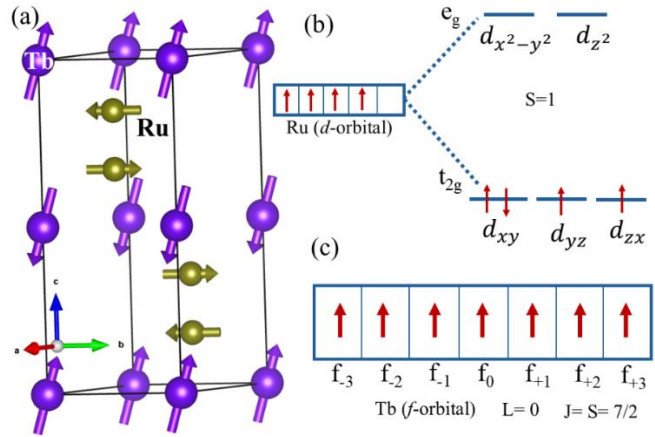

**Fig. 4 (a)** Magnetic Structure of $Ba_3TbRu_2O_9$, Spin-configuration for **(b)** Ru and **(c)** Tb spins in $Ba_3TbRu_2O_9$.

The Tb possesses a +4-valence state, contrary to the usual +3 valence state of R-ions in this whole family. Most likely, the energy is minimized by adopting this spin-configuration similar to the related $Gd^{3+}$. The spin configurations for Tb and Ru spins are shown in Fig.4b and 4c.

The slightly smaller value of the refined Tb moment compared to its theoretical spin-only value of 7/2 could be due to two reasons: i) the Landé g-factor might be slightly lower than 2 (the ideal value for a free electron's spin-only moment) in this strongly correlated electron system with complex magnetic interaction; and/or ii) the moments may not be fully saturated even at 2 K, although this is unlikely. Interestingly, unlike other R-members in this series where a reduced Ru-moment is generally observed, our results yield an almost full moment for Ru (1.96 µB) [5,12]. We did not observe any molecular-like magnetic state as observed in $Ba_3YRu_2O_9$ or a selective-orbital state similar to $Ba_3LaRu_2O_9$ [3,10].

**Conclusion**

In summary, we resolve the spin-structure of the $Ba_3TbRu_2O_9$ compound and observe a unique magnetic ground state for both terbium and ruthenium in $Ba_3TbRu_2O_9$, distinct from all other lanthanide members in this series. The 4*f* orbitals of Tb behave like those of an s-state ion, exhibiting zero orbital moment (**L**=0). Our neutron diffraction results reveal that the magnetic moment of Ru-atom is almost 2 µB manifesting a **S**=1 magnetic ground state, in contrast to a reduced moment of ruthenium in all other members in this family. We also report a cooperative spin-ordering of both Ru and Tb, not commonly observed in *d-f* coupled systems.

**Acknowledgement**

T.B. greatly acknowledges the Science and Engineering Research Board (SERB) (Project No.: SRG/2022/000044), and UGC-DAE Consortium for Scientific Research (CSR) (Project No CRS/2021-22/03/544), Government of India, and SEED Grant, RGIPT, for research funding. A portion of this research used resources at the Spallation Neutron Source, a DOE Office of Science User Facility operated by the Oak Ridge National Laboratory. The beam time was allocated to the SNAP instrument on proposal number IPTS- 33740.1. We acknowledge Dr. Jhuma Sannigrahi, IIT Goa for fruitful discussion. TB thank the Central Instrumentation Facilities (CIF), RGIPT. M.K. thanks the University Grant Commission (UGC), India, for the research fellowship.

# Supplementary Information

1. **Scanning Electron Microscopy (SEM)/ Energy-Dispersive X-ray spectroscopy (EDX):**

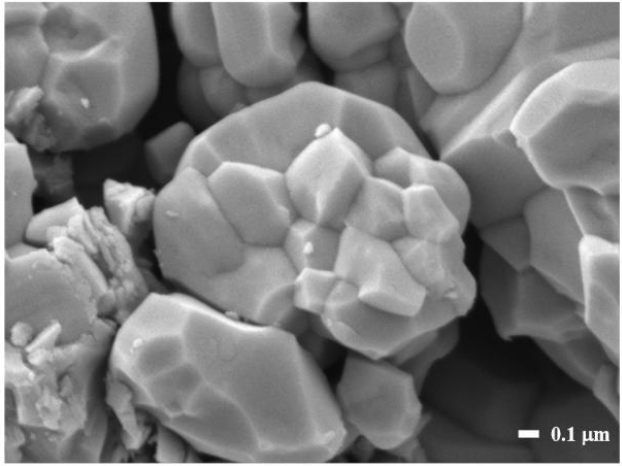

**Figure.S1** Scanning electron micrographs of $Ba_3TbRu_2O_9$ powder sample.

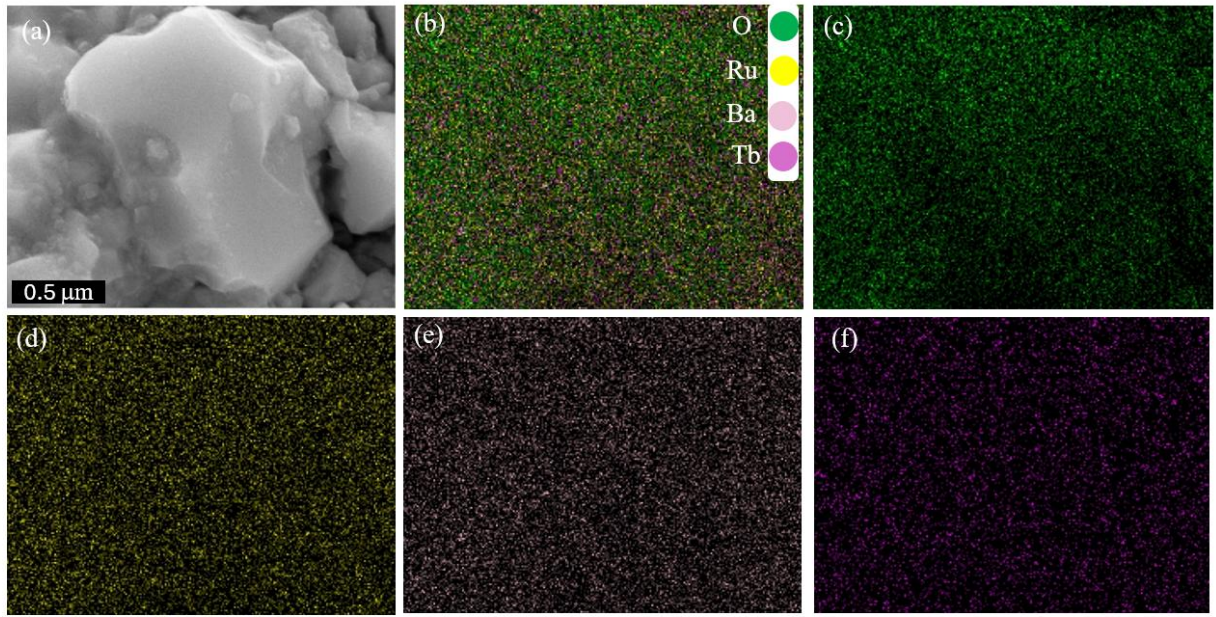

**Figure.S2** (a) Backscattered electron Backscattered electron image, and EDX maps display (b) the mixing of Ba, Tb, Ru and O, (c) O, (d) Ru, (e) Ba and (f) Tb of $Ba_3TbRu_2O_9$ powder sample.

2. **Transmission Electron Microscopy (TEM):**

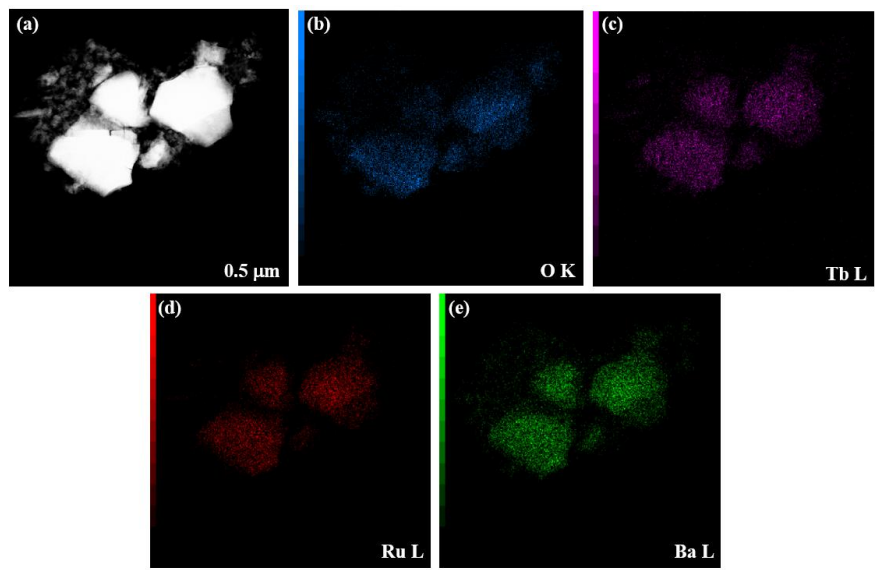

**Figure.S3** (a) HAADF-STEM image of polycrystalline $Ba_3TbRu_2O_9$, showing well-defined grains with sharp boundaries. Elemental mapping obtained from EDS analysis: (b) oxygen (O K), (c) terbium (Tb L), (d) ruthenium (Ru L), and (e) barium (Ba L).

3. **X-ray Photoelectron Spectroscopy (XPS):**

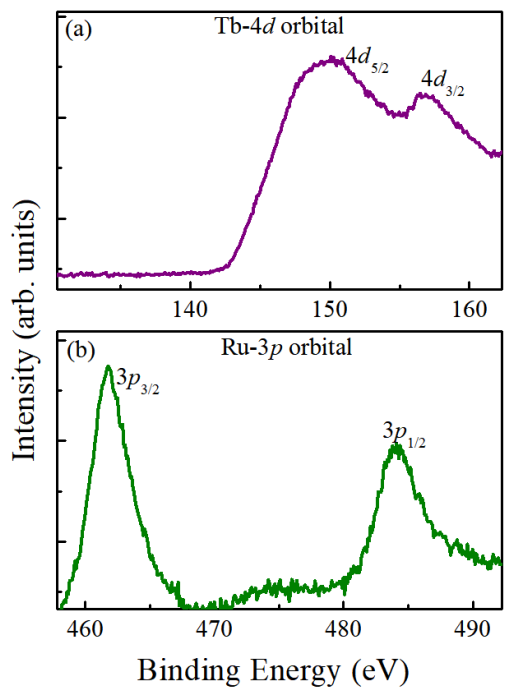

**Figure.S4** X-ray photoelectron Spectroscopy (XPS) of (a)Tb $4d_{5/2}$ and $4d_{3/2}$, and (b) Ru $3p_{3/2}$ and $3p_{1/2}$ for $Ba_3TbRu_2O_9$

## 4. Time-of-Flight Neutron Diffraction:

Fig.S5 shows the Rietveld refinement fitting of 100 K data, that demonstrates the high quality of the fit.

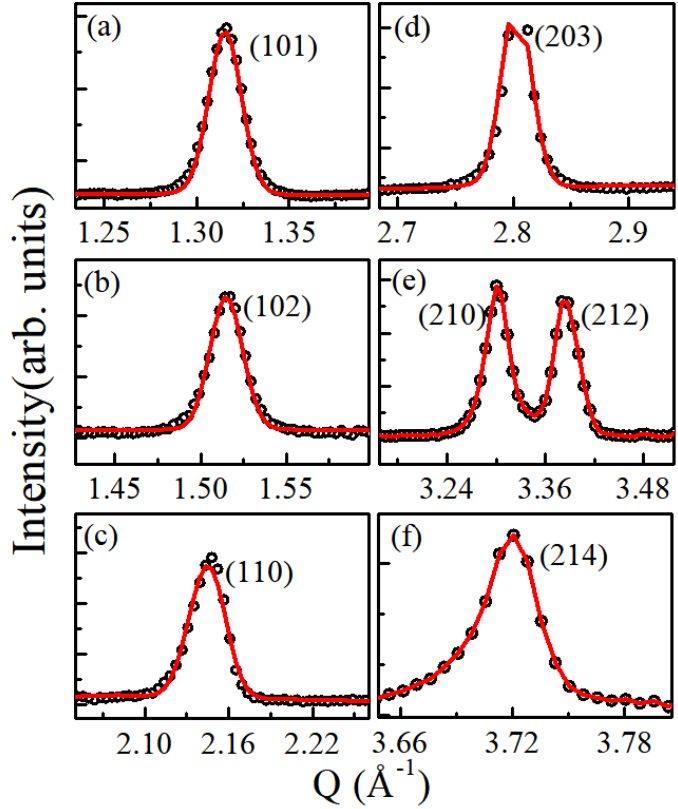

**Figure.S5** Enlarged view of selected Bragg peaks from the 100 K TOF neutron diffraction data, showing individual peak fittings. The experimental data are shown as black open circles, and the calculated fits are shown as solid red lines. Panels (a)–(f) correspond to the (101), (102), (110), (203), (210)/ (212), and (214) reflections, respectively.

**Table S1:** Basis vector for the space group P6$_3$/mmc for K= (0 0 0). The decomposition of the magnetic representation for the Tb site (0, 0, 0) is $\Gamma_{mag}(Tb) = 0\Gamma_1^1 + 0\Gamma_2^1 + 1\Gamma_3^1 + 0\Gamma_4^1 + 0\Gamma_5^1 + 0\Gamma_6^1 + 1\Gamma_7^1 + 0\Gamma_8^1 + 1\Gamma_9^2 + +0\Gamma_{10}^2 + 1\Gamma_{11}^2 + 0\Gamma_{12}^2$. The atoms of the nonprimitive basis are defined according to 1: (0, 0, 0), 2:(0, 0, .5).

| IR | BV | Atom | BV Components | | | | | |
|---|---|---|---|---|---|---|---|---|
| | | | m∥a | m∥b | m∥c | im∥a | im∥b | im∥c |
| $\Gamma_3$ | $\Psi_1$ | 1 | 0 | 0 | 12 | 0 | 0 | 0 |
| | | 2 | 0 | 0 | 12 | 0 | 0 | 0 |
| $\Gamma_7$ | $\Psi_2$ | 1 | 0 | 0 | 12 | 0 | 0 | 0 |
| | | 2 | 0 | 0 | -12 | 0 | 0 | 0 |

| IR | BV | Atom | | | | | | |
|---|---|---|---|---|---|---|---|---|
| | | | m∥a | m∥b | m∥c | im∥a | im∥b | im∥c |
| $\Gamma_9$ | $\Psi_3$ | 1 | 0 | -3 | 0 | 0 | 0 | 0 |
| | | 2 | 0 | -3 | 0 | 0 | 0 | 0 |
| | $\Psi_4$ | 1 | -3.464 | -1.732 | 0 | 0 | 0 | 0 |
| | | 2 | -3.464 | -1.732 | 0 | 0 | 0 | 0 |
| $\Gamma_{11}$ | $\Psi_5$ | 1 | 0 | -3 | 0 | 0 | 0 | 0 |
| | | 2 | 0 | 3 | 0 | 0 | 0 | 0 |
| | $\Psi_6$ | 1 | -3.464 | -1.732 | 0 | 0 | 0 | 0 |
| | | 2 | 3.464 | 1.732 | 0 | 0 | 0 | 0 |

**Table S2: Basis vectors consistent with space group P6$_3$/mmc for K= (0 0 0). The decomposition of the magnetic representation for the Ru site (.33333, .66667, .16223) is $\Gamma_{mag}(Ru) = 0\Gamma_1^1 + 1\Gamma_2^1 + 1\Gamma_3^1 + 0\Gamma_4^1 + 0\Gamma_5^1 + 1\Gamma_6^1 + 1\Gamma_7^1 + 0\Gamma_8^1 + 1\Gamma_9^2 + +1\Gamma_{10}^2 + 1\Gamma_{11}^2 + 1\Gamma_{12}^2$. The atoms of the nonprimitive basis are defined according to 1: (.33333, .66667, .16223), 2: (.66666, .33333, .66223), 3:(.66667, .33334,.83777), 4: (.33333, .66667, .33777).**

| IR | BV | Atom | BV Components | | | | | |
|---|---|---|---|---|---|---|---|---|
| | | | m∥a | m∥b | m∥c | im∥a | im∥b | im∥c |
| $\Gamma_2$ | $\Psi_1$ | 1 | 0 | 0 | 6 | 0 | 0 | 0 |
| | | 2 | 0 | 0 | 6 | 0 | 0 | 0 |
| | | 3 | 0 | 0 | -6 | 0 | 0 | 0 |
| | | 4 | 0 | 0 | -6 | 0 | 0 | 0 |
| $\Gamma_3$ | $\Psi_2$ | 1 | 0 | 0 | 6 | 0 | 0 | 0 |
| | | 2 | 0 | 0 | 6 | 0 | 0 | 0 |
| | | 3 | 0 | 0 | 6 | 0 | 0 | 0 |
| | | 4 | 0 | 0 | 6 | 0 | 0 | 0 |
| $\Gamma_6$ | $\Psi_3$ | 1 | 0 | 0 | 6 | 0 | 0 | 0 |
| | | 2 | 0 | 0 | -6 | 0 | 0 | 0 |
| | | 3 | 0 | 0 | -6 | 0 | 0 | 0 |
| | | 4 | 0 | 0 | 6 | 0 | 0 | 0 |
| $\Gamma_7$ | $\Psi_4$ | 1 | 0 | 0 | 6 | 0 | 0 | 0 |
| | | 2 | 0 | 0 | -6 | 0 | 0 | 0 |
| | | 3 | 0 | 0 | 6 | 0 | 0 | 0 |
| | | 4 | 0 | 0 | -6 | 0 | 0 | 0 |
| $\Gamma_9$ | $\Psi_5$ | 1 | 0 | -1.5 | 0 | 0 | 0 | 0 |
| | | 2 | 0 | -1.5 | 0 | 0 | 0 | 0 |
| | | 3 | 0 | -1.5 | 0 | 0 | 0 | 0 |
| | | 4 | 0 | -1.5 | 0 | 0 | 0 | 0 |
| | $\Psi_6$ | 1 | -1.732 | -0.866 | 0 | 0 | 0 | 0 |
| | | 2 | -1.732 | -0.866 | 0 | 0 | 0 | 0 |
| | | 3 | -1.732 | -0.866 | 0 | 0 | 0 | 0 |
| | | 4 | -1.732 | -0.866 | 0 | 0 | 0 | 0 |
| $\Gamma_{10}$ | $\Psi_7$ | 1 | 3 | 1.5 | 0 | 0 | 0 | 0 |
| | | 2 | 3 | 1.5 | 0 | 0 | 0 | 0 |
| | | 3 | -3 | -1.5 | 0 | 0 | 0 | 0 |
| | | 4 | -3 | -1.5 | 0 | 0 | 0 | 0 |
| | $\Psi_8$ | 1 | 0 | -2.598 | 0 | 0 | 0 | 0 |
| | | 2 | 0 | -2.598 | 0 | 0 | 0 | 0 |
| | | 3 | 0 | 2.598 | 0 | 0 | 0 | 0 |
| | | 4 | 0 | 2.598 | 0 | 0 | 0 | 0 |
| $\Gamma_{11}$ | $\Psi_9$ | 1 | 0 | -1.5 | 0 | 0 | 0 | 0 |
| | | 2 | 0 | 1.5 | 0 | 0 | 0 | 0 |
| | | 3 | 0 | -1.5 | 0 | 0 | 0 | 0 |
| | | 4 | 0 | 1.5 | 0 | 0 | 0 | 0 |
| | $\Psi_{10}$ | 1 | -1.732 | -0.866 | 0 | 0 | 0 | 0 |
| | | 2 | 1.732 | 0.866 | 0 | 0 | 0 | 0 |
| | | 3 | -1.732 | -0.866 | 0 | 0 | 0 | 0 |
| | | 4 | 1.732 | 0.866 | 0 | 0 | 0 | 0 |
| $\Gamma_{12}$ | $\Psi_{11}$ | 1 | 3 | 1.5 | 0 | 0 | 0 | 0 |
| | | 2 | -3 | -1.5 | 0 | 0 | 0 | 0 |
| | | 3 | -3 | -1.5 | 0 | 0 | 0 | 0 |
| | | 4 | 3 | 1.5 | 0 | 0 | 0 | 0 |
| | $\Psi_{12}$ | 1 | 0 | -2.598 | 0 | 0 | 0 | 0 |
| | | 2 | 0 | 2.598 | 0 | 0 | 0 | 0 |
| | | 3 | 0 | 2.598 | 0 | 0 | 0 | 0 |
| | | 4 | 0 | -2.598 | 0 | 0 | 0 | 0 |

**Table S3: Refined magnetic moment components $m_a$, $m_b$ and $m_c$ along crystallographic a, b, c directions and total magnetic moment ($m_{total}$) for Tb and Ru atoms in $Ba_3TbRu_2O_9$, obtained from Fullprof magnetic structure refinement.**

| Atom | $m_a$ | $m_b$ | $m_c$ | $m_{total}$ |
|---|---|---|---|---|
| Tb | 0 | ±1.69200 | ±5.9520 | 6.18782 |
| Ru | 0 | ±1.96050 | 0 | 1.96050 |

**References:**

[1] Y. Ishiguro, K. Kimura, S. Nakatsuji, S. Tsutsui, A. Q. Baron, T. Kimura and Y. Wakabayashi, *Nature Communications*, 2013, 4, 2022.

[2] J. Hwang, E. Choi, F. Ye, C. Dela Cruz, Y. Xin, H. Zhou and P. Schlottmann, *Physical review letters*, 2012, 109, 257205.

[3] D. Ziat, A. A. Aczel, R. Sinclair, Q. Chen, H. Zhou, T. J. Williams, M. B. Stone, A. Verrier and J. Quilliam, *Physical Review B*, 2017, 95, 184424.

[4] T. Basu, V. Caignaert, S. Ghara, X. Ke, A. Pautrat, S. Krohns, A. Loidl and B. Raveau, *Physical Review Materials*, 2019, 3, 114401.

[5] T. Basu, V. Caignaert, F. Damay, T. Heitmann, B. Raveau and X. Ke, *Physical Review B*, 2020, 102, 020409.

[6] E. Kushwaha, G. Roy, M. Kumar, A. Dos Santos, S. Ghosh, D. Adroja, V. Caignaert, O. Perez, A. Pautrat and T. Basu, *Physical Review B*, 2024, 109, 224418.

[7] S.-J. Kim, M. D. Smith, J. Darriet and H.-C. Zur Loye, *Journal of Solid State Chemistry*, 2004, 177, 1493–1500.

[8] W. Miiller, M. Avdeev, Q. Zhou, A. J. Studer, B. J. Kennedy, G. J. Kearley and C. D. Ling, *Physical Review B—Condensed Matter and Materials Physics*, 2011, 84, 220406.

[9] S. Hayashida, H. Gretarsson, P. Puphal, M. Isobe, E. Goering, Y. Matsumoto, J. Nuss, H. Takagi, M. Hepting and B. Keimer, *Physical Review B*, 2025, 111, 104418.

[10] Q. Chen, A. Verrier, D. Ziat, A. Clune, R. Rouane, X. Bazier Matte, G. Wang, S. Calder, K. M. Taddei, C. d. Cruz et al., *Physical Review Materials*, 2020, 4, 064409.

[11] T. Basu, A. Pautrat, V. Hardy, A. Loidl and S. Krohns, *Applied Physics Letters*, 2018, 113, year.

[12] M. S. Senn, S. A. Kimber, A. M. Arevalo Lopez, A. H. Hill and J. P. Attfield, Physical *Review B—Condensed Matter and Materials Physics*, 2013, 87, 134402.

[13] Y. Doi and Y. Hinatsu, *Journal of Materials Chemistry*, 2002, 12, 1792–1795.

[14] Y. Doi, M. Wakeshima, Y. Hinatsu, A. Tobo, K. Ohoyama and Y. Yamaguchi, *Journal of Materials Chemistry*, 2001, 11, 3135–3140.

[15] Q. Chen, S. Fan, K. M. Taddei, M. B. Stone, A. I. Kolesnikov, J. Cheng, J. L. Musfeldt, H. Zhou and A. A. Aczel, *Journal of the American Chemical Society*, 2019, 141, 9928–9936.

[16] J. Sannigrahi, A. Paul, A. Banerjee, D. Khalyavin, A. Hillier, K. Yokoyama, A. Bera, M. R. Lees, I. Dasgupta, S. Majumdar et al., *Physical Review B*, 2021, 103, 144431.

[17] T. Basu, F. Wei, Q. Zhang, Y. Fang and X. Ke, *Physical Review Materials*, 2020, 4, 114401.

[18] G. Maris, V. Volotchaev and T. Palstra, *New Journal of Physics*, 2004, 6, 153.

[19] J. Rodríguez-Carvajal, *Physica B: Condensed Matter*, 1993, 192, 55–69. [20] A. S. Wills, Physica B: Condensed Matter, 2000, 276, 680-681.

[21] L. Jozwiak, J. Balcerzak and J. Tyczkowski, *Catalysts*, 2020, 10, 278.

[22] E. G. Sogut, H. Acidereli, E. Kuyuldar, Y. Karatas, M. Gulcan and F. Sen, *Scientific reports*, 2019, 9, 15724.

[23] 23 Z. Cao, C. Wang, Y. Sun, M. Liu, W. Li, J. Zhang and Y. Fu, *Chemical Science*, 2024, 15, 1384–1392.

[24] X. Mao, Z. Liu, C. Lin, J. Li and P. K. Shen, *Inorganic Chemistry Frontiers*, 2023, 10, 558–566.

[25] L. Chang, M. Prager, J. Perßon, J. Walter, E. Jansen, Y. Chen and J. Gardner, *Journal of Physics: Condensed Matter*, 2010, 22, 076003.

[26] N. Taira, M. Wakeshima and Y. Hinatsu, *Journal of Materials Chemistry*, 2002, 12, 1475-1479.

[27] C. Wiebe, J. Gardner, S.-J. Kim, G. Luke, A. Wills, B. Gaulin, J. Greedan, I. Swainson, Y. Qiu and C. Jones, *Physical review letters*, 2004, 93, 076403.

[28] Y. Doi, Y. Hinatsu, K.-i. Oikawa, Y. Shimojo and Y. Morii, *Journal of Materials Chemistry*, 2000, 10, 797–800.

[29] D. Adroja, S. Sharma, C. Ritter, A. Hillier, D. Le, C. Tomy, R. Singh, R. Smith, M. Koza, A. Sundaresan et al., *Physical Review B*, 2020, 101, 094413.

[30] Y. Doi, Y. Hinatsu, A. Nakamura, Y. Ishii and Y. Morii, *Journal of Materials Chemistry*, 2003, 13, 1758–1763.